# Phase-cycling and double-quantum two-dimensional electronic spectroscopy using a common-path birefringent interferometer

Daniel Timmer[a*], Daniel C. Lünemann[a], Moritz Gittinger[a], Antonietta De Sio[a,b], Cristian Manzoni,[c] Giulio Cerullo[c,d] and Christoph Lienau[a,b*]

[a] Institut für Physik, Carl von Ossietzky Universität Oldenburg, 26129 Oldenburg, Germany

[b] Center for Nanoscale Dynamics (CENAD), Carl von Ossietzky Universität Oldenburg, 26129 Oldenburg, Germany

[c] Istituto di Fotonica e Nanotecnologie-CNR, Piazza L. da Vinci 32, 20133 Milano, Italy

[d] Dipartimento di Fisica, Politecnico di Milano, Piazza L. da Vinci 32, 20133 Milano, Italy

*Correspondence to: christoph.lienau@uni-oldenburg.de or daniel.timmer@uol.de

## Abstract

Selecting distinct quantum pathways in two-dimensional electronic spectroscopy (2DES) can give particularly deep insights into coherent and incoherent interactions and quantum dynamics in various materials. This includes isolating rephasing and non-rephasing pathways for conventional single-quantum 2DES, but also the ability to record double- and zero-quantum spectra. Such advanced 2DES schemes usually require phase-cycling when performed in a partially or fully collinear geometry. A particularly simple and effective implementation of 2DES utilizes an in-line birefringent interferometer, the Translating-Wedge-based Identical pulses eNcoding System (TWINS), for the generation of an inherently phase-stable collinear excitation pulse pair. Here, we demonstrate how the TWINS can be adapted to allow for phase-cycling and experimental access to isolated quantum pathways. These new capabilities are demonstrated by recording rephasing, non-rephasing, zero-quantum and double-quantum 2DES on a molecular J-aggregate. This easy-to-implement extension opens up new experimental possibilities for TWINS-based 2DES in multidimensional all-optical and photoemission spectroscopy and microscopy.

## Introduction

Transient absorption (TA) spectroscopy is a highly developed technique to probe the quantum dynamics of optical excitations of matter with a time resolution that is solely limited by the duration of the employed pump and probe pulses. In TA, an ultrashort pump pulse impulsively excites the material and pump-induced changes in optical properties are measured by the absorption changes of a time-delayed probe pulse. In conventional two-pulse TA, spectral information can only be accessed about the detection step, while the information about the excitation step is lost in the impulsive pumping. Ultrafast two-dimensional electronic spectroscopy[1-4] (2DES), derived from multidimensional nuclear magnetic resonance spectroscopy,[5] is an extension of TA spectroscopy that overcomes this limitation by exciting the sample with two pump pulses delayed by $\tau$ and obtaining resolution in excitation frequency exploiting a Fourier transform approach. The resulting 2D maps correlate the excitation and detection frequencies as a function of the waiting time $T$ between the second pump pulse and the probe pulse, maximizing simultaneously temporal resolution and frequency selectivity. These 2DES maps allow to investigate, e.g., coherent and incoherent couplings in hybrid quantum materials,[6-10] many-body interactions in semiconductors[11-13] or vibronic couplings in molecular systems.[14, 15] 2DES is typically performed in two different experimental geometries: (i) a fully non-collinear geometry, also known as BoxCARS, in which all three pulses propagate along different directions and phase matching ensures that only the desired signals are generated in the propagation direction that reaches the detector; (ii) a partially collinear geometry, also called pump-probe geometry, where the two pump pulses are collinear and the nonlinear signal propagates along the direction of the probe beam.

The measured third-order response is conveniently described in terms of perturbation theory,[4, 16, 17] giving rise to certain sets of quantum pathways that depend on the field interactions between the three laser pulses and the investigated system. Depending on the sign of the coherences that are evolving after the first and the third field interactions, the system can either follow so-called rephasing (R) or non-rephasing (NR) pathways. The former one represents the photon echo case, in which dephasing effects that lead to inhomogeneous broadening are reversed, allowing to isolate the effects of inhomogeneous broadening.[18] Separating R and NR pathways offers additional insights into interactions and coherent dynamics. They can be used, e.g., to distinguish between ground and excited state vibrational wavepacket motion in molecular systems[19] or to disentangle many-body interactions in semiconductors.[12, 20] In order to facilitate their separation, some degree of freedom in the field interaction is required. For example, directional phase matching can be used in the BoxCARS geometry, taking advantage of the different angles under which the three pulses impinge on the sample.[2, 4] For partially (or fully) collinear 2DES, careful manipulation of the phases of the pulses via phase-cycling (PC)[21-25] or frequency-tagging[26-28] is required. Here, 2DES data are recorded for different absolute phase values that are imprinted onto the excitation pulses and the desired quantum pathways are then isolated from a linear combination of the nonlinear signals.

In addition to isolating R and NR pathways, PC also allows to record a distinctly different class of 2DES experiments.[4, 29, 30] Often, 2DES investigations are limited to so-called single-quantum (1Q) experiments, probing transitions between the ground state and one-quantum manifold and between the one-quantum and two-quantum manifolds. These experiments give access to the excited state population dynamics. In addition, the third order nature of 2DES also allows to record quantum pathways that are termed as zero-quantum (0Q) and double-quantum (2Q) signals. In contrast to 1Q 2DES, 0Q and 2Q 2DES probe coherences between states that are separated by zero and two quanta of the excitation energy, respectively. They therefore provide information about two-quantum coherences and coherences within the excited state manifold. This can give additional and direct insights into coherent couplings, (many-body) interactions and doubly-excited states in the system.[30-33] It should be noted that these experiments differ from higher-order 2DES and TA schemes that can also access information about multiply-excited states.[34-36]

Currently, two main approaches are established for 2DES in the pump-probe geometry: pulse shapers and birefringent interferometers. Pulse shapers, using either a 4f configuration[37] or an acousto-optic programmable dispersive filter,[21, 38] act on the spectral amplitude and phase of the input pulse to create a phase-stable collinear pump-pulse pair. The full phase control allows for advanced acquisition schemes via PC.[21, 24, 25, 37] Shaper-based 2DES therefore enables the selection of quantum pathways and to separately record R, NR as well as 0Q and 2Q 2DES spectra.[38, 39] An alternative approach is based on a common-path interferometer, the Translating Wedge-based Identical pulse eNcoding System[40] (TWINS), that has been demonstrated to be a simple and comparatively cost-effective solution for the generation of inherently phase-stable excitation pulse pairs for 2DES.[40, 41] TWINS can be used for a wide range of wavelengths[41-43] and broadband laser sources[43] while maintaining a high time-resolution. It only requires control of a motorized delay stage to tune the pump-pulse delay with high accuracy and repeatability. However, one major drawback of the TWINS, in particular when compared to shaper-based pulse generation, is the apparent lack of phase-cycling capabilities.[2, 3] So far, this significant restriction excluded TWINS as an option for advanced 2DES experiments that require to isolate distinct quantum pathways.

Here, we report on a simple and straightforward solution to the phase-cycling problem for TWINS. We demonstrate that the insertion of an achromatic quarter-wave plate (QWP) allows to manipulate the two polarization components inside the TWINS and unlocks full control of the relative absolute phase between the pump pulses. This opens up advanced 2DES schemes for TWINS such as isolating R and NR contributions but also 0Q and 2Q spectroscopy. These new capabilities are demonstrated for 2DES on a molecular J-aggregate thin-film at room temperature. The sample acts as an almost ideal effective three-level-system (3LS), allowing for quantitative comparison to perturbative simulations of the 2DES spectra.

## Results and Discussion

### *Traditional TWINS*

In the TWINS, a single block and two wedge pairs made out of birefringent crystals are used to create an inherently phase-stable, collinearly propagating pulse pair. Fig. 1 depicts the working principle of the TWINS, which exploits the difference between ordinary and extraordinary refractive indexes in birefringent materials, which creates different group delays for the vertically and horizontally polarized field components.[40] By shaping a birefringent block in the form of two wedges and laterally translating one of the wedges, the group delay $\tau$ between the pulses can be tuned with high precision. In its original design,[40] TWINS consists of two polarizers, here denoted as $Pol_1$ and $Pol_2$, set to 45°, and three sets of differently cut alpha-barium borate ($\alpha$-BBO) birefringent crystals. Those are a single block ("Y-cut") and two wedge pairs ("Z-cut" and "X-cut"), with labels denoting the orientation of the optical axis of the birefringent material, as indicated in the top row in Fig. 1. Briefly, the first polarizer prepares the input pulse at 45° linear polarization relative to the vertical, creating equal field amplitudes for the s- and p-polarization components, oriented at 0° and 90°, respectively. The first polarizer could be replaced by a half-wave plate to increase throughput by a factor of 2. These s- and p-polarized components then propagate within the Y-cut block with a group velocity set by either the extraordinary or ordinary refractive index, $n_e$ or $n_o$, respectively. As a result, a frequency-dependent phase difference $\Delta\phi(\omega) = \phi_2(\omega) - \phi_1(\omega)$ is introduced which is proportional to the birefringence $\Delta n(\omega) = n_e(\omega) - n_o(\omega)$. This creates a fixed group delay $\tau_g = \partial\Delta\phi(\omega)/\partial\omega$ between the two polarization components (see section 1 of the Supporting Information, SI). This delay can be compensated by the two X-cut wedges, for which ordinary and extraordinary refractive indexes are exchanged. If the beam path inside the two wedges matches the Y-cut block thickness (2 mm), no phase difference between the two polarization components and thus no group delay is introduced. Mechanical tuning of the X-cut wedge insertion can then be used to create both positive and negative delays between the s- and p-polarized components. The Z-cut wedges are isotropic so that both components see the refractive index $n_o$, and are used to ensure that the total amount of traversed material is kept constant. For this, the $X_1$ and $Z_2$ wedges are moved synchronously (Fig. 1, top). Finally, the second polarizer ($Pol_2$, case i) in Fig. 1) projects both polarization components onto 45° to ensure that both pulses are polarized along the same direction. A more quantitative description can be found in Ref 41 and in the SI.

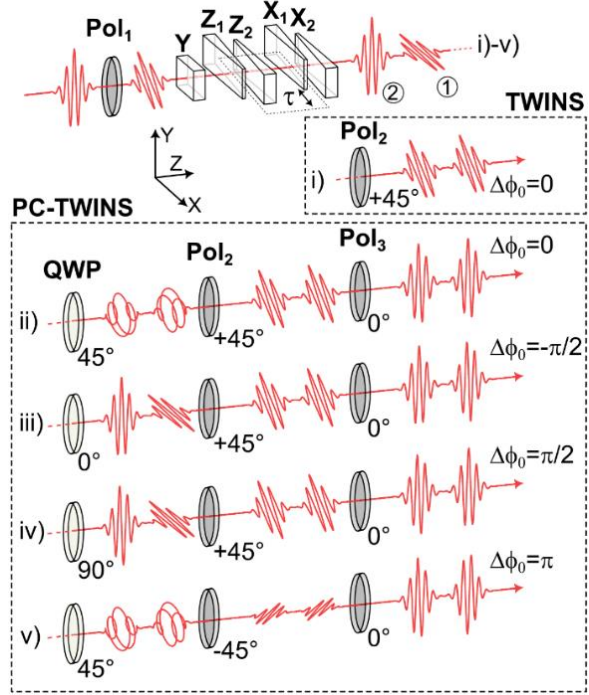

*Figure 1: Scheme of the PC-TWINS for phase-cycling. The single input pump pulse, polarized at 0° relative to the vertical passes the first polarizer (Pol$_1$) with 45° polarization. The birefringent α-BBO Y-cut block with its optical axis aligned along the Y-direction and two pairs of α-BBO wedges (Z$_1$, Z$_2$, X$_1$ and X$_2$) with optical axes oriented along Z and X, respectively, split the input pump pulse and create a pump-pulse pair with tunable delay τ. The leading pulse (1) is p-polarized, while the trailing pulse (2) is s-polarized for τ > 0. Translation of the Z$_2$ and X$_1$ wedges only shifts the first pulse (1). The next row, case i), depicts the traditional TWINS case where a second polarizer projects both pulses onto a common polarization direction. For the PC-TWINS, here shown with cases ii)-v), a quarter-wave plate (QWP) introduces a spectrally-constant phase-difference $\Delta\phi_0$ between the two output pulses. ii)-iv): Orientations of the fast axis at 45°, 0° and 90° introduce phase differences of $0, -\pi/2$ and $+\pi/2$, respectively. The second polarizer (Pol$_2$) projects the polarization of both pulses onto 45°. v): When Pol$_2$ is turned from +45° to -45°, an additional phase of π is introduced. The final and optional polarizer (Pol$_3$) ensures constant polarization and output intensity for all displayed phase settings.*

## PC-TWINS

Our aim is to adapt the TWINS such that it allows for PC. This requires to introduce a spectrally constant difference $\Delta\phi_0$ between the absolute phases,[44] also known as carrier-envelope phases (CEP), of the two excitation pulses. Phase settings of multiples of $\pi/2$ are of particular interest for 2DES applications. So far, translation of the Z$_2$ and X$_1$ wedges in the conventional TWINS only introduces, to first order, a linear phase to the first excitation pulse. The relative phase $\Delta\phi$ is constant and zero for a well-aligned TWINS with Pol$_1$ and Pol$_2$ both set to 45°. In order to create absolute phase differences as required for PC, we utilize a zero-order achromatic quarter-wave plate (B. Halle) that is placed inside the TWINS (in-between Pol$_1$ and Pol$_2$). The bottom rows ii)-v) in Fig. 1 conceptually depict the effects of the QWP on the phases of the two field components for selected QWP settings. Additional phases, i.e. dispersion effects, that identically affect both polarization components are neglected for clarity. In case ii), the QWP is oriented at 45°, converting the s- and p-polarized components into counter-rotating, circularly polarized light. The light then passes through Pol$_2$, which has the same 45° orientation as the fast axis of the QWP. This creates a linearly polarized pulse pair after Pol$_2$. For this pair, no phase difference is introduced, $\Delta\phi_0 = 0$, since both transmitted pulses experienced the same refractive index in the QWP. When turning the QWP to 0° (case iii)), the fast axis is oriented vertically and therefore a phase corresponding to a quarter-wave, i.e. $\pi/2$, is introduced to pulse 1, while pulse 2 remains unaffected, creating $\Delta\phi_0 = -\pi/2$. When turning the QWP to 90° (case iv)), the second pulse is retarded instead, resulting in $\Delta\phi_0 = +\pi/2$. Lastly, as shown in v), the second polarizer Pol$_2$ can be used to introduce an additional phase $\phi_2 = \pi$ when turned from 45° to -45°. Therefore, $\Delta\phi_0 = \pi$ can be obtained when setting the QWP to 45° but Pol$_2$ to -45°.

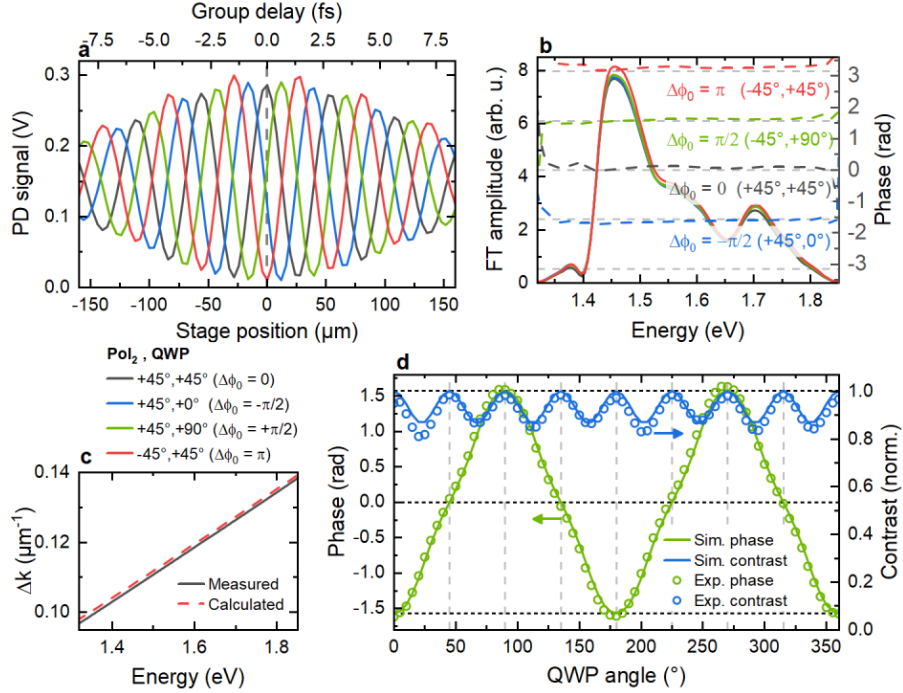

*Figure 2: Phase-cycling using the PC-TWINS interferometer. **a:** Central region of linear autocorrelation measurements for four selected phase settings of 0, $\pi/2$, $-\pi/2$ and $\pi$, recorded behind the PC-TWINS as a function of the stage position using a photo diode (PD). The group delay is shown for 800 nm. **b:** Spectra and phases obtained from the autocorrelations depicted in panel a via Fourier transform. The energy axis is calculated from the measured calibration curve depicted in c. **c:** Wavevector $\Delta k(E)$ that defines the group delay introduced to the time-delayed first pump-pulse via the birefringence of the $\alpha$-BBO crystals. The measured curve (black, solid) is obtained from a spectral interferogram scan. Small deviations to the calculated (red, dashed) curve likely arise from manufacturing tolerances, in particular of the apex angle of the wedges. **d:** Measured phase and contrast as a function of the QWP orientation (circles) compared to calculations (solid lines) based on Jones calculus. The measurements demonstrate that continuous phase tuning can be achieved using the QWP. For phases that are a multiple of $\pi/2$, a high contrast is maintained.*

To verify these considerations, linear autocorrelation traces (Fig. 2a) of broadband (680-900 nm) laser pulses (Fig. S2c) are measured behind the PC-TWINS with a photo diode (PD) for these four phase settings. They clearly show the expected relative CEP shifts, reflecting the absolute phases of $\Delta\phi_0 = [0, \pi/2, -\pi/2, \pi]$. No shifts of the envelope are observed. Fourier transforms of the full autocorrelation traces, yielding the pulse spectra and phase differences $\Delta\phi(\omega)$, are shown in Fig. 2b. While the frequency-dependent phase of the TWINS interferometer and therefore the energy axis can in principle be calculated from the material birefringence, it has proven to be useful to record a calibration scan[40, 41] that characterizes the actual frequency-dependent phases introduced when translating the wedges (see SI section 4). The calibration scan is performed by recording a spectral interferogram produced by the PC-TWINS as a function of stage position and wavelength. This way, the wavelength-dependent modulation frequency (in µm$^{-1}$) for a stage scan can be assigned to a photon energy, as shown in Fig. 2c and explained in more detail in Section 4 of the SI.[40, 41] The measured spectral phases $\Delta\phi(\omega)$ confirm that indeed absolute phases $\Delta\phi_0$ in multiples of $\pi/2$ are introduced. Tuning of the spectral phase does not affect the intensity of the spectral interference term, nor its spectral dependence. A small signal increase for the case of $\Delta\phi_0 = \pi$ (red line) likely reflects small alignment inaccuracies of the third polarizer.

A more quantitative analysis of the effect of the QWP can be obtained from Jones calculus[45] (see Section 1 of the SI). For an angle of the QWP of $\vartheta$ (fast axis relative to the vertical) we obtain phase differences

$$\Delta\phi_0(\vartheta) = -\arctan\left(\frac{\cos(2\vartheta)}{\sin^2(2\vartheta)}\right) + \phi_2 \text{ with } \phi_2 = \begin{cases} 0, \text{Pol}_2 \text{ at } +45° \\ \pi, \text{Pol}_2 \text{ at } -45° \end{cases} \quad (1)$$

This shows that the introduced phase difference is not restricted to values that are a multiple of $\pi/2$, but any arbitrary phase between $-\pi/2$ and $\pi/2$ can be set using the QWP. All remaining phases can be covered by turning the second polarizer to -45°. In addition, we can calculate the interference contrast reflecting an imbalance between the two field components

$$C(\vartheta) = \sqrt{1 - \frac{1}{4}\sin^2(4\vartheta)} \quad (2)$$

According to Eq. (2), the contrast is periodically modulated between 1 and ~0.87 as a function of $\vartheta$. Importantly, however, a high contrast of 1 is expected for relevant phase-cycling values, i.e. multiples of $\pi/2$. Additional tuning of the polarizers could in principle be used to counteract this systematic contrast modulation (see SI).

The expected phase and contrast modulations are experimentally verified as presented in Fig. 2d. Using a measured autocorrelation and spectral interferogram as a function of the QWP orientation, $\Delta\phi_0(\vartheta)$ (green circles) and $C(\vartheta)$ (blue circles) are deduced from a Fourier transform of the PD signal and a spectral interference spectrum, respectively. They agree well with the theoretically expected curves (solid lines), demonstrating the high accuracy and reproducibility of the phase-cycling abilities of the PC-TWINS.

*Rephasing and non-rephasing spectra*

To demonstrate the new capabilities of the PC-TWINS, we perform 2DES on a 10-nm thin-film of molecular J-aggregates[46-48] on a gold substrate. In this sample, dipolar interactions between the ordered ProSQ-C16 squaraine[46, 49] molecules, a prototypical quadrupolar charge-transfer chromophore, result in a delocalization of the optical excitation across ~24 molecules[47] and the formation of superradiant J-aggregate exciton states $|X\rangle$.[47, 50, 51] The large oscillator strength and sharp exciton resonance (Fig. 3a) make these aggregates particularly well-suited for strong coupling experiments.[6] Since many molecules contribute to the collective excitation, the aggregate supports multi-exciton states. Within one aggregate, Pauli-blocking of the lowest energy exciton state results in a blue-shift of the two-exciton state $|XX\rangle$, which is also superradiant.[47, 50] As depicted in Fig. 3a and discussed in more detail in Ref. 47, the J-aggregate can be considered as an effective 3LS. The dominant origin of the nonlinearity is therefore the two-exciton blue-shift $\Delta E$.[50, 52]

We use a home-built high-repetition rate 2DES setup that has previously been reported in Ref. 47 (see Section 3 of the SI). We utilize 10-fs laser pulses, covering ~680-900 nm, generated by a home-built noncollinear optical parametric amplifier[47, 53] that is pumped by a Yb:KGW laser (Light Conversion, Carbide) operating at 200 kHz repetition rate (see Section 3 of the SI). Pulse durations are optimized and characterized by TWINS-based interferometric and pump-probe cross-correlation frequency-resolved optical gating measurements (Fig. S2). Mechanical chopping of the pump is used to record differential spectra that contain the third order nonlinear signal. For performing conventional 2DES, as well as for isolating R and NR spectra, we employ the pulse ordering depicted in Fig. 3b. Using the PC-TWINS, we scan the coherence time $\tau > 0$ for $T = 0$ fs. Both pump and probe are s-polarized. The sample is measured in vacuum and at room temperature. A pump fluence of ~10 μJ/cm² ensures that the signals are collected within the third order regime.[6, 47] Even though all experiments are performed in reflection geometry, the gold substrate, acting as a mirror, gives rise to spectra that effectively show absorptive lineshapes due to the double-transmission geometry.[6, 13, 47]

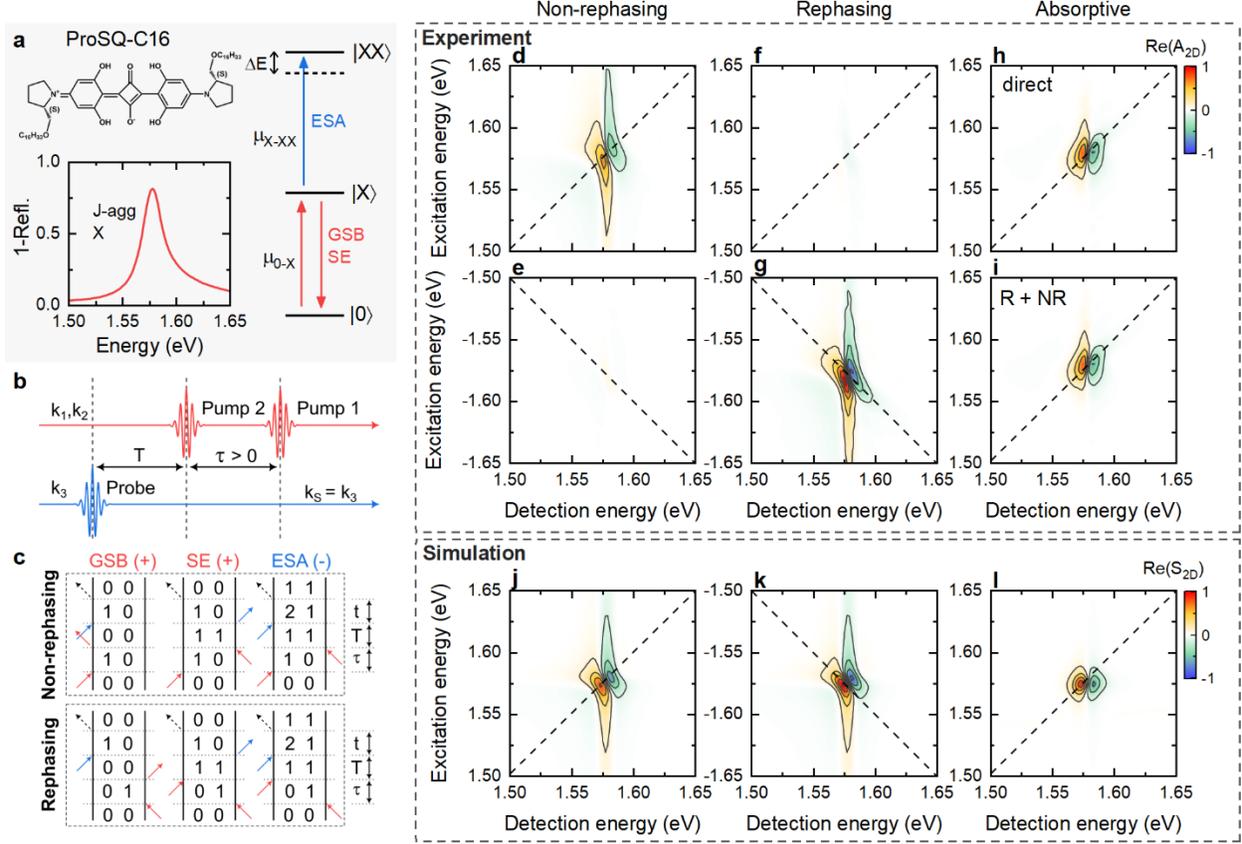

*Figure 3:* Rephasing (R) and non-rephasing (NR) spectra for squaraine J-aggregates acquired by phase-cycling with $\Delta\phi_0 = [0, +\pi/2]$. *a:* Absorption spectrum (bottom), taken as 1 − Refl. (Refl.: sample reflectivity), of the molecular J-aggregate based on ProSQ-C16 squaraines (top). It shows a sharp exciton (X) resonance at 1.578 eV. The system acts as an effective 3LS with a blue-shifted two-exciton (XX) state (right). *b:* Experimental pulse scheme for 1Q 2DES. *c:* Relevant double-sided Feynman diagrams for the third-order response of the 3LS. *d-i:* Real parts of the experimental 2DES maps for $T = 0$ fs. Phase-cycling allows to isolate NR (d,e) and R (f,g) spectra. The larger amplitude of the R relative to the NR signal is a consequence of finite inhomogeneous broadening in the sample. Their sum (R+NR, i) matches the conventionally obtained absorptive 2DES map (h). *j-l:* Real part of simulated NR (j), R (k) and absorptive (l) maps for a 3LS using the response functions corresponding to the diagrams in c, a dephasing time of $T_2^{1Q} = 160$ fs, and a two-exciton shift $\Delta E = 3$ meV. A finite inhomogeneous broadening of $\sigma_{inh} = 2$ meV is included.

Conventional absorptive 2DES is first recorded using $\Delta\phi_0 = 0$ by setting the QWP to 45°. Fig. 3h presents the real part of the 2DES signal $Re(A_{2D})$, in good agreement with previous studies.[6, 47] The dispersive lineshape along the detection energy reflects positive ground state bleaching (GSB) and stimulated emission (SE) as well as negative excited state absorption (ESA) transitions. The measured absorptive 2DES map thus shows the expected nonlinearity of the effective 3LS.

Next, a two-step phase-cycling scheme is employed by recording two sets of 2DES data for phase settings of $\Delta\phi_0 = [0, \pi/2]$. R and NR spectra are obtained from these two measurements following a data evaluation procedure described by Ogilvie and co-workers.[21] In short, causality along the detection time is enforced, resulting in R and NR contributions being separated in the frequency domain by residing in different spectral quadrants.[21, 54] Subsequent linear combinations of the appropriately weighted complex-valued 2DES maps then yield isolated R and NR maps, as discussed in more detail in Section 4 of the SI.[17, 21, 25] The resulting 2DES spectra are only located in a distinct spectral quadrant of the excitation-detection energy plane and their real parts are displayed in Fig. 3d-g. Importantly, the sum of the R and NR maps has to reproduce the absorptive 2DES map after reflecting the rephasing spectrum along the $E_{ex} = 0$ line.[17] As seen in Fig. 3i, the isolated R and NR spectra convincingly allow us to reconstruct the absorptive spectrum. Additional imaginary contributions are displayed in Fig. S3.

The double-sided Feynman diagrams[16, 17] that describe all quantum pathways of such a 3LS for the experimental scheme in Fig. 3b are listed in Fig. 3c. We simulate the R, NR and absorptive 2DES spectra using the response functions corresponding to the double-sided Feynman diagrams. The results are presented in Fig. 3j-l. Convincing agreement between simulation and experiment is achieved when choosing a 1Q dephasing time of $T_2^{1Q} = 160$ fs, a two-exciton shift $\Delta E = 3$ meV and a finite inhomogeneous broadening of $\sigma_{inh} = 2$ meV (see SI for more details). We use the same dephasing time for the $|0\rangle \leftrightarrow |X\rangle$ as for the $|X\rangle \leftrightarrow |XX\rangle$ transition. A finite inhomogeneous broadening contribution is required to reproduce the difference in signal amplitude between the R and NR spectra. Note that the homogeneous dephasing time of 160 fs that results from this analysis is significantly longer than expected based on earlier measurements on similarly prepared samples.[6, 47] Also the inhomogeneous broadening of the transition is very weak, giving evidence for a high structural homogeneity of the spin-coated thin films investigated here. A crosscut for a fixed excitation energy in Fig. S1 also confirms the quantitative agreement between the simulated and measured 2DES lineshapes. Additional measurements on HITCI dye dissolved in ethanol are reported in Figs. S4 and S5.

*Double-quantum and zero-quantum spectra*

Using the phase-cycling capabilities of the PC-TWINS, we can further investigate zero- and double-quantum properties of the molecular aggregate. For the investigated 3LS, the 2Q spectra are of particular interest to directly access the two-exciton state and to gain information about the two-exciton blueshift and its dephasing properties. To probe the 0Q and 2Q pathways for 2DES in pump-probe geometry, we follow a data acquisition and evaluation scheme that was recently demonstrated by Cai et al.[39] for a shaper-based approach using PC. Fig. 4a shows the applied acquisition scheme with a reversed pulse ordering.[39, 55] To ensure that the response is emitted into the probe direction, 2DES data are recorded for $T = 0$ fs and negative coherence times ($\tau < 0$). We employ a 2-step phase cycling with $\Delta\phi_0 = [0, \pi/2]$ and perform the same data evaluation procedure as for the R and NR data (see Section 4 of the SI). Since the nonlinear signal is emitted after the last field interaction, which is no longer the probe but rather pump 1, a $\tau$-dependent (linear) phase is imprinted onto the measured signal.[39] To correct for the effective rotating frame that results from this phase shift, the proper excitation axis for the 0Q and 2Q spectra is obtained after shifting the excitation energy by the detection energy, $E_{0Q/2Q}(E_{det}) = E_{ex} + E_{det}$.

The resulting real parts of the 2Q and 0Q spectra for the J-aggregate are shown in Fig. 4c,d. Imaginary parts are shown in Fig. S3. Similar to the 1Q spectra, the observed dispersive lineshape is a result of the superposition of different quantum pathways that contribute with opposite signs, as seen in the double-sided Feynman diagrams in Fig. 4b. Simulations of the 2Q and 0Q spectra based on these response functions are presented in Fig. 4e,f. Reasonable agreement with experiment can be achieved using the same parameters as for the simulation of the 1Q spectra. Finite deviations are likely due to our simplistic model that does not take excitonic fine-structure into account.[47, 50, 51] The 0Q and 2Q measurements can now give additional information, in particular they allow us to estimate the two-exciton blue-shift $\Delta E$.

Fig. 4g shows both the absolute value of the 2Q and reconstructed $(R + NR)$ 1Q excitation profile (see SI). From the spectral shift between the maxima of both curves, we estimate $\Delta E \approx 3 \pm 2$ meV.

In the 2Q and 0Q experiments, probe and pump 2 coincide and are preparing the system in a double- or zero-quantum coherence, respectively. For the investigated 3LS, the 2Q case allows us to extract the 2Q dephasing time $T_2^{2Q} = 30$ fs (Fig. 4g). Even though we do not attempt a microscopic modelling of this dephasing time,[51] we note that a phenomenological Lindblad dephasing model[6] predicts a 2Q dephasing time that is four times shorter than $T_2^{1Q}$. This is reasonable agreement with the measured value of 30 fs. In contrast, the linewidth of the 0Q spectrum (Fig. 4h) represents population relaxation which we estimate as $T_1 = 140$ fs. This time constant agrees with a previously observed partial population decay on a 100-fs time scale in these aggregates.[6] Since currently only a finite delay range for negative $\tau$ of ~230 fs is available for our PC-TWINS, additional slower relaxation processes cannot be captured at the moment.

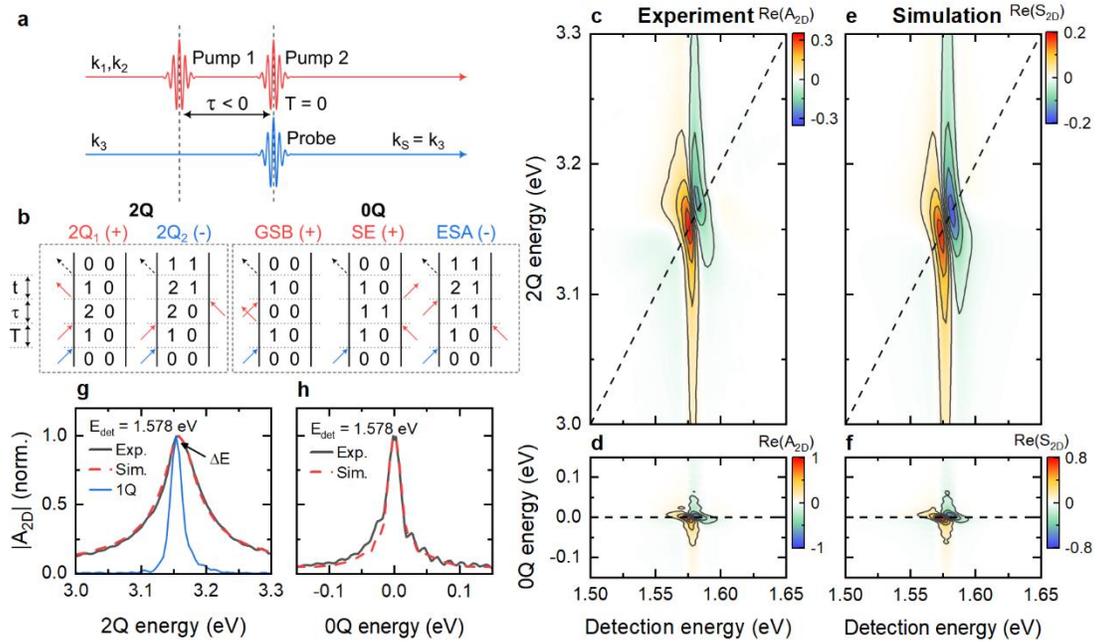

*Figure 4:* 2Q and 0Q spectra for the squaraine J-aggregate. *a:* Experimental pulse scheme adapted for the acquisition of 0Q and 2Q spectra. By choosing $T = 0$ fs, the probe and second pump coincide. The first pump is now trailing since only negative coherence times $\tau < 0$ are scanned. Using this scheme, the desired third order response is radiated into the probe direction. *b:* Double-sided Feynman diagrams for the 2Q and 0Q spectra derived for the time-ordering in a. *c,d:* Real part of the experimental 2Q (c) and 0Q (d) spectra for the J-aggregate. *e,f:* Real part of the respective simulated spectra for the 3LS using the same parameters as in Fig. 3. *g:* Crosscut through the absolute-valued 2Q peak. A good agreement with the simulation is achieved when using a 2Q dephasing time $T_2^{2Q} = 30$ fs. A comparison to the 1Q profile ($abs(R + NR)$), shifted by $E_{det}$ along the excitation energy, allows to estimate the two-exciton shift $\Delta E$. *h:* Crosscut through the absolute-valued 0Q peak. Using the simulation, a time constant of $T_1 = 140$ fs is deduced. At the moment, coherence time scans using $\tau < 0$ are limited to a maximum delay of ~230 fs. The same truncation of the $\tau$-axis is used in the simulation.

## Conclusion and outlook

We have presented a simple and effective way to introduce phase-cycling capabilities for the TWINS interferometer. By adding an achromatic quarter-wave plate, full and continuous relative absolute phase control between the excitation pulses can be achieved. We demonstrate these new capabilities of the PC-TWINS for 2DES on a J-aggregated molecular thin film, comprising the prototypical quadrupolar charge-transfer chromophore squaraine. The experiments reveal an exceptionally long dephasing time of the J-aggregated exciton transition of 160 fs at room temperature. A weak inhomogeneous line broadening and a small two-exciton shift of 3 meV support the favorable optical properties of the investigated thin films and, in particular, their usefulness for studies of strong coupling phenomena. The experimental results also suggest to extend the reported investigations to low temperature in order to further enlarge the exciton coherence time.

Furthermore, we have used a 2-step phase-cycling scheme to isolate rephasing and non-rephasing contributions and 0Q and 2Q spectra. Comparison to simulated spectra highlights the high fidelity of the technique. In principle, all recorded data could be obtained from a single forth and back scan of the PC-TWINS when covering both positive and negative coherence times. A fast, motorized rotation mount can be used to rapidly toggle between chosen phase values in-between scans, making the recorded data relatively insensitive to slow laser drifts or sample degradation.

We believe that this new methodology will substantially extend the experimental possibilities of TWINS-based coherent spectroscopy and eliminates one major drawback compared to shaper-based approaches. This extension of the TWINS can be easily added to existing setups and allows control of the excitation sequence that was previously inaccessible. The selection of quantum pathways and in particular access to double-quantum 2DES will be of crucial importance for investigating coherent couplings and many-body interactions in quantum materials. The broad phase-cycling opportunities may also include developments toward TWINS-based 2DES microscopy[22, 23, 28] of isolated nanostructures. Beyond all-optical spectroscopy, it may find fruitful applications for future combinations of multidimensional and photoemission spectroscopies.[22, 56]

**Data availability**

Data underlying the results presented in this paper are not publicly available at this time but may be obtained from the authors upon reasonable request.

**Supporting information**

Experimental methods, additional experimental data, simulation details, Figs. S1-S5, Table S1 (PDF).

**Notes**

G.C. discloses involvement with NIREOS, a company that commercializes the TWINS interferometer.

**Acknowledgements**

We acknowledge financial support from Deutsche Forschungsgemeinschaft (SFB1372/2-Sig01 "Magnetoreception and Navigation in Vertebrates", INST 184/163-1, INST 184/164-1, Li 580/16-1, and DE 3578/3-1). We also acknowledge financial support from the Niedersächsische Ministerium für Wissenschaft und Kultur (DyNano and Wissenschaftsraum ElLiKo) and the Volkswagen Foundation (SMART). G.C. acknowledges financial support by the European Union's NextGenerationEU Programme with the I-PHOQS Infrastructure [IR0000016, ID D2B8D520, CUP B53C22001750006] "Integrated infrastructure initiative in Photonic and Quantum Sciences". C.M. acknowledges financial support by the European Union's European Innovation Council (EIC), TROPHY (PATHFINDER OPEN-01 call, grant Nr. 101047137).

**TOC Figure**

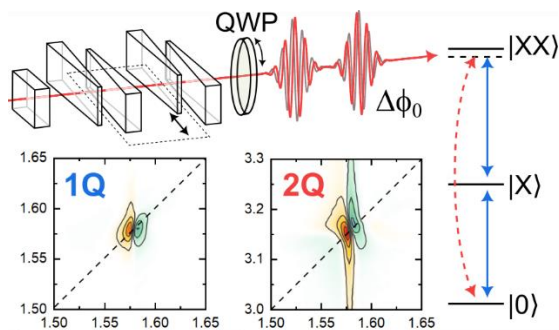